\documentclass[vecphys]{svmult}

\usepackage{graphicx}        
\usepackage[bottom]{footmisc}

\begin{document}

\title*{X-ray Emission from Planetary Nebulae and their Central Stars: a Status Report}
\titlerunning{X-rays from PNe: a Status Report} 

\author{Joel H. Kastner\inst{1,2}
}
\institute{Center for Imaging Science, Rochester Institute of Technology, Rochester, NY 14623 USA
\texttt{jhk@cis.rit.edu}
\and Visiting Astronomer, Laboratoire
    d'Astrophysique de Grenoble, Universit\'e Joseph Fourier --- CNRS, BP 53, 38041 Grenoble Cedex, France 
}
%
%
\maketitle

\begin{abstract}
  In the era of {\it Chandra} and {\it XMM-Newton}, the detection (or
  nondetection) of diffuse and/or point-like X-ray sources within
  planetary nebulae (PNe) yields important, unique insight into PN
  shaping processes. Diffuse X-ray sources, whether due to ``hot
  bubbles'' or to collimated outflows or jets, allow us to probe the
  energetic shocks within PN wind interaction
  regions. Meanwhile, X-ray point sources provide potential
  diagnostics of magnetic fields, accretion disks, and/or binary
  companions at PN cores. Here, I highlight recent X-ray observational
  results and trends that have the potential to shed new light on the origin
  and evolution of the structure of PNe. 
\end{abstract}

\section{Introduction}

Over the past decade, X-ray imaging by the {\it Chandra} and {\it
  XMM-Newton} X-ray Observatories has provided fresh, compelling
observational evidence for hot bubbles, highly energetic jets, and/or
``active'' central sources within planetary nebulae (PNe). So far,
nine of $\sim25$ PNe targeted by these two contemporary X-ray
observatories\footnote{Martin Guerrero maintains a list
  of PNe observed by {\it Chandra} or {\it XMM} at
  \texttt{http://www.iaa.es/xpn/}.} have been detected as diffuse
X-ray sources (Kastner et al.\ 2000, 2001, 2003; Chu et al.\ 2001;
Guerrero et al.\ 2002, 2005; Sahai et al.\ 2003; Montez et al.\ 2005;
Gruendl et al.\ 2006), while an additional handful have been revealed
to harbor X-ray point sources at their cores (e.g., Guerrero et al.\
2001; Kastner et al.\ 2003). Both ``flavors'' of X-ray source ---
diffuse and point-like --- can be
used to probe PN shaping processes and to constrain models of PN
structural evolution.

\section{Diffuse X-ray Sources}

\subsection{X-rays from  PN ``hot bubbles''}

Models of PN shaping have long predicted the formation of
X-ray-emitting ``hot bubbles.'' A hot bubble may be produced within a
PN as the central star makes the transition from post-asymptotic giant
branch (post-AGB) star to white dwarf, following an evolutionary track
of increasing $T_\star$ at constant $L_\star$, followed by decreasing
$L_\star$ and $T_\star$. In this phase the star produces very fast and
energetic winds (with speeds $\sim1000$ km s$^{-1}$ and mass loss
rates $\stackrel{>}{\sim}10^{-7}$ $M_\odot$ yr$^{-1}$).  When such a
fast wind collides with ambient (previously ejected AGB) gas, it is
shocked and superheated (e.g., Zhekov \& Perinotto 1996). The shocked
fast wind forms an overpressured bubble that accelerates outwards and
displaces the ambient (visible, nebular) gas as it grows (Kwok,
Purton, \& Fitzgerald 1978). The supersonic growth of the bubble plows
the displaced older material into a rim of dense gas which, when
projected on the sky, is seen as a thin molecular, dust, and/or
ionized structure that traces the bubble's perimeter.

\begin{figure}[ht]
\begin{center}
\includegraphics[scale=0.5,angle=0]{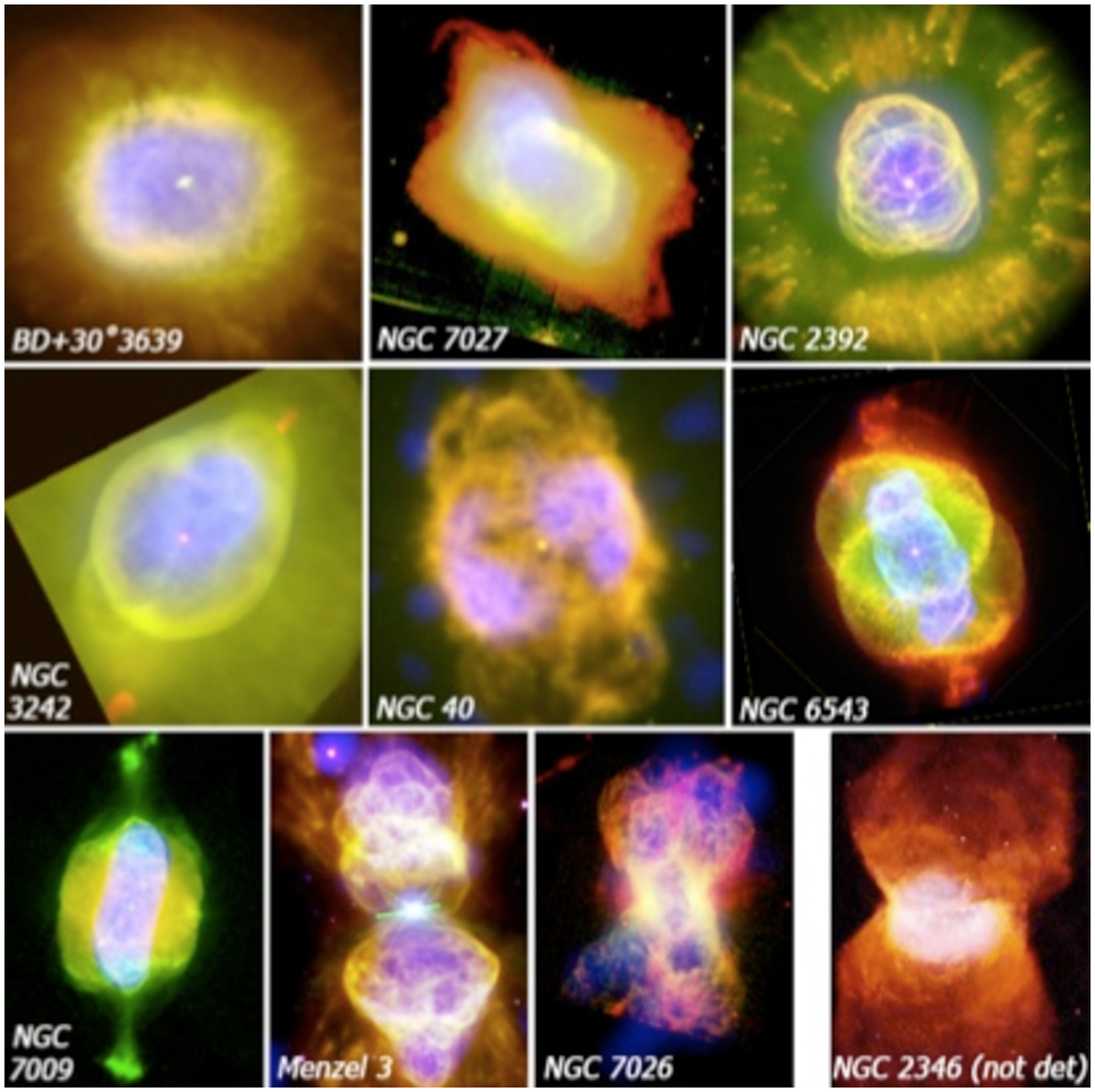}
\end{center}
\caption{Images of all of the PNe for which extended X-ray emission
  (shown in blue) has been detected (as of this writing) by {\it
    Chandra} (BD +30$^\circ$3639, Mz 3, NGC 6543, 7027, 40) or {\it
    XMM-Newton} (NGC 2392, 3242, 7009, 7026) --- and one XMM
  nondetection (NGC 2346, with open bipolar lobes) --- arranged
  according to optical morphology. (Image montage courtesy Bruce Balick
  and Martin Guerrero.) }
\end{figure}

In the majority of detections of diffuse X-ray emission from PNe, the
X-ray-emitting region is fully contained within bright optical rims or
bubbles (Fig.\ 1), as predicted by the preceding ``hot bubble''
scenario. Furthermore, many of the detected objects harbor central
stars that are of Wolf-Rayet (WR) type (i.e., ``[WC]'' or ``[WO]''
stars) and/or display optical spectroscopic evidence for large
mass-loss rates and wind velocities. These trends were already
reasonably well established (Montez et al.\ 2005;
Gruendl et al.\ 2006) when we stumbled upon yet another example of an
X-ray-emitting ``hot bubble'' within a [WC] PN: NGC 5315.

\subsubsection{{\it Chandra's} serendipitous detection of NGC
  5315}

In a Cycle 5 {\it Chandra} study, we observed two [WC] PNs, NGC 40 and
Hen 2-99; we detected the former, but failed to detect the latter
(Montez et al.\ 2005). In March 2007, while searching the Chandra
archives for targeted observations of PNs in support of a {\it
  Chandra} observing proposal, Rudy Montez discovered --- to our great
surprise and amusement --- that a {\it second} [WC] PN, NGC~5315,
was present at the edge of the field of our 29
ks {\it Chandra} Advanced CCD Imaging Spectrometer (ACIS) observation
targeting Hen 2-99. Rudy's examination of the Chandra/ACIS image revealed
an X-ray source at the position of NGC~5315, $\sim12.5'$ off-axis.
This serendipitous {\it Chandra} detection
of X-rays from NGC 5315 is all the more remarkable when one considers
that there are only $\sim50$ known [WC] PNe in the sky (Gorny \&
Stasinska 1995; Tylenda 1996). Of course, it could also be said that
this author pointed one of NASA's Great Observatories at the
wrong object! Indeed, due to the relatively poor {\it Chandra}/ACIS-S
image quality at the far off-axis position of NGC~5315, it is not
possible to ascertain from the ACIS image alone whether the X-rays
trace a hot bubble within this PN, emanate from the PN nucleus, or are
emitted by both the nebula and its central star(s).

\begin{figure}[htb]
\begin{center}
\includegraphics[scale=0.3,angle=-90]{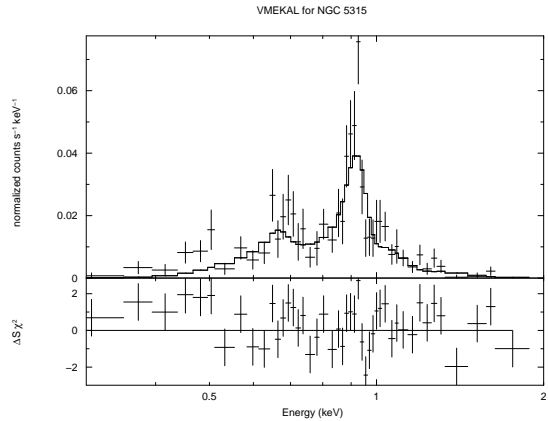}
\end{center}
\caption{ Chandra/ACIS spectrum of the X-ray source associated with
  NGC~5315 (crosses), with best-fit absorbed thermal plasma (VMEKAL)
  model spectrum overlaid. The fit residuals are indicated in the
  lower panel. (From Kastner et al.\ 2007.)}
\label{fig:NGC5315spec}
\end{figure}

Fortunately, the background-subtracted spectrum
(Fig.~\ref{fig:NGC5315spec}), the luminosity, and the temporal
behavior of the X-ray source associated with NGC~5315 appear quite
definitive regarding the origin of the X-rays. The spectrum shows
strong Ne~{\sc ix} line emission as well as a blend of O~{\sc vii} and
O~{\sc viii} lines, with no evidence of Fe L-shell lines. Spectral
modeling indicates that the emission arises in a $\sim2.5\times10^6$ K thermal
plasma with enhanced Ne and depleted Fe.  These results, and the
inferred intrinsic source luminosity of $L_X \sim 2\times10^{32}$ erg
s$^{-1}$, are very similar to those obtained for the best-characterized
diffuse X-ray PN, BD +30$^\circ$3639 (Kastner et al.\ 2000, 2006a).
Meanwhile, the Chandra/ACIS light curve shows no
evidence for variability, and the absorption-corrected X-ray
luminosity of NGC~5315 is at least an order of magnitude larger than
that of any unresolved PN core region detected thus far (Guerrero et
al.\ 2001; Kastner et al.\ 2003), further supporting the
interpretation that the X-rays arise from an extended region within
NGC~5315. These results, described in detail in Kastner et al.\
(2007), establish NGC~5315 as one of the most luminous ``hot bubble''
PN X-ray sources yet detected.

\subsection{NGC 5315 and other hot bubbles: the picture thus far}

The detection of the [WC] PN NGC~5315 by 
{\it Chandra} underscores two significant trends that
have emerged from the X-ray observations of PNe obtained
thus far by {\it Chandra} and {\it XMM} (Kastner et al.\ 2007):
\vspace{-0.1cm}
\begin{enumerate}
\item Objects with WR-type (i.e., [WC], [WO], or WR(H)) central stars
  --- which display characteristically large wind velocities ($v_w$)
  and mass-loss rates ($\dot{M}$) --- account for a disportionately large
  fraction of PNe that are established sources of luminous,
  diffuse X-ray emission. Specifically, five of the seven known
  ``hot bubble'' PN X-ray sources are associated with WR-type PNe.
\item In all cases in which diffuse X-ray emission is detected, the
  optical/IR structures that enclose the regions of diffuse X-rays are
  clearly defined, and these structures generally display thin,
  bright, uninterrupted edges (or ``rims'') surrounding a cavity of
  lower surface brightness that is coincident with the extended X-ray
  emission (see also Gruendl et al.\ 2006).
\end{enumerate}
\vspace{-0.1cm}
These results indicate that the combination of (1)
large wind kinetic energies ($L_w \stackrel{>}{\sim}
10^{33}$ erg s$^{-1}$) and (2) a ``closed containment
vessel'' is necessary to yield PN hot bubbles with plasma 
densities sufficient to produce detectable soft (0.3-2.0
keV) X-ray luminosities $L_X \stackrel{>}{\sim} 10^{31}$ erg
s$^{-1}$. 

\begin{figure}[htb]
\begin{center}
\includegraphics[scale=0.5,angle=90]{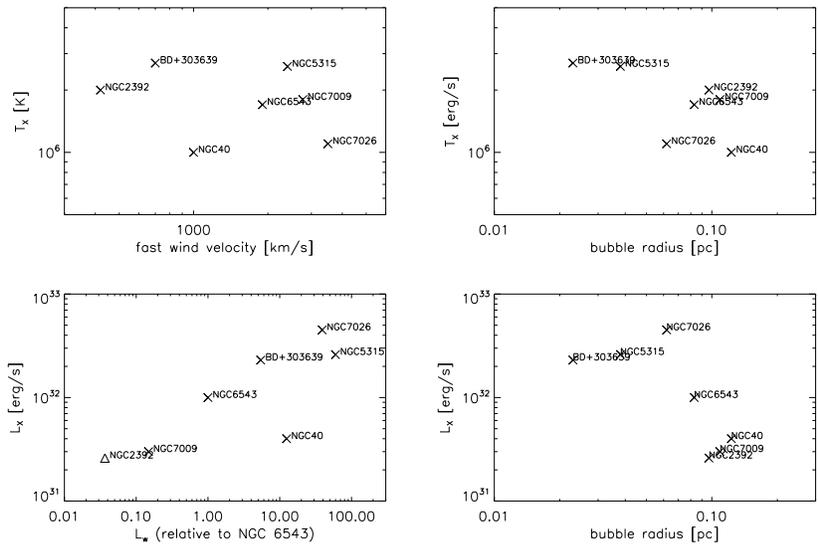}
\end{center}
\caption{ Results obtained thus far for PNe
  detected as ``hot bubble'' X-ray sources by Chandra and
  XMM-Newton. Upper left:
  PN plasma temperature ($T_X$) vs.\ central star fast wind
  velocity. Upper right: $T_X$ vs.\ PN optical bubble
  radius. Lower left: PN X-ray luminosity ($L_X$) vs.\ wind
  luminosity of the PN central star ($L_w$) relative to
  $L_w$ of the central star of NGC 6543 ($L_w$ for NGC 2392
  [triangle] is an upper limit). Lower right: $L_X$ vs.\ PN optical bubble
  radius. (From Kastner et al.\ 2007.) } 
\label{fig:PNXraytrends} \end{figure}

Fig.~\ref{fig:PNXraytrends} 
displays scatter diagrams for various PN and central star parameters measured
for the X-ray-detected objects. The top panels demonstrate that the
characteristic temperature of the X-ray-emitting plasma is far lower
than expected, based on simple shock models, in almost all PNs in
which diffuse X-ray emission has been detected thus far. Indeed, there
is only one ``hot bubble'' PN, NGC 2392, for which the predicted
temperature of the shocked wind gas is consistent with $T_X$; for all
other such objects, the predicted post-shock temperatures are larger
than observed by factors ranging from $\sim2$ (BD $+30^\circ$3639) to
$\sim200$ (NGC 7026). It also appears that $T_X$ is uncorrelated with
present-day central star wind velocity, but is weakly anticorrelated
with PN bubble radius (Fig.~\ref{fig:PNXraytrends}, upper panels),
suggesting that PN age is more important than present-day wind kinetic
energy in determining the temperature of the X-ray-emitting plasma
within hot bubbles.

The bottom panels of Fig.~\ref{fig:PNXraytrends} indicate
that X-ray luminosity is correlated with present-day central
star wind luminosity $L_w = \frac{1}{2} \dot{M}v_w^2$, and
is perhaps anticorrelated with bubble radius. 
The shallow slope of the $L_X$ vs.\ $L_w$ correlation --- wherein 4
orders of magnitude in $L_w$ results in only a factor $\sim20$ range
in $L_X$ --- may suggest either that the conversion of wind kinetic
energy to plasma radiation becomes more efficient as the central star
wind declines in strength, or that the luminosities (and, perhaps,
temperatures) of the hot bubbles within these PNe are established
during early phases of the post-AGB evolution of central stars with
rapidly evolving winds (Akashi et al.\ 2007; see also
Sch\"onberner et al. 2006).

\subsection{X-rays from collimated outflows and jets}

A review of X-rays from collimated outflows and jets in PNe was
presented by Martin Guerrero at APN IV (these
  proceedings). Here, I merely point out that detections of X-ray
emission from the energetic shocks that are expected to result from
the interaction of PN collimated outflows, jets, and/or bullets with
slower-moving (AGB) material remain few and far between;
only 2 or 3 such detections have been made to date (Hen 3-1475, Sahai
et al.\ 2003; Mz~3, Kastner et al.\ 2003; and, perhaps, NGC 7027,
Kastner et al.\ 2001). The dearth of X-ray detections of PN jets and
jet-like structures is not particularly surprising, in light of the
similarly small X-ray detection rate of highly collimated
outflows from young stellar objects (e.g., Grosso et al.\ 2006 and
references therein). 

Nevertheless, some trends may be emerging. In particular, it
seems that the characteristic temperatures of X-ray sources associated
with PNe jets ($\sim3$--6 MK) are systematically higher than those of
``hot bubble'' PNe. Also, as would be expected, all
three ``X-ray jet'' PNe show evidence for dusty, molecule-rich
(H$_2$-emitting) central tori that are potential collimating agents
for the outflows that produce the X-ray-emitting shocks, whereas only
one ``hot bubble'' PN X-ray source (BD+30$^\circ$3639) was detected in
the H$_2$ survey of Kastner et al.\ (1996).

\section{Point-like X-ray Sources within PNe}

We have been revisiting all {\it Chandra} observations
targeting PNe, so as to place constraints on the X-ray luminosities of
PNe central star systems (Kastner \& Montez, in prep.). The superior
($\sim0.5"$) spatial resolution of {\it Chandra} (compared with {\it
  XMM-Newton's} $\sim5"$ resolution) is required to distinguish such
point-like X-ray emission from diffuse emission originating with the
surrounding nebula. To ascertain the {\it Chandra} point-source X-ray
luminosities or luminosity upper limits, we determine the ACIS count
rate within an aperture of radius $\sim2.5"$ placed at the nominal
position of the PN central star (this radius accounts for typical {\it
  Chandra} pointing errors of $\stackrel{<}{\sim}1"$). We then compare
this ACIS count rate with that obtained for a suitably chosen
background region (typically, an annulus surrounding the source
region).

\subsection{``Classical'' PNe}

Applying this technique to {\it Chandra} observations of PNe whose
central stars are not symbiotic in nature (see below), we confirm the
detection of X-ray point sources within four nebulae (NGC 246, NGC
4361, NGC 6543, and NGC 7293; see Guerrero et al.\ 2001 and
\verb+http://www.iaa.es/xpn/+) and obtain X-ray flux upper limits for
10 other nebulae (including three sources of diffuse X-ray emission,
i.e., NGC 40, BD+30, and NGC 7027). We hence obtain a preliminary (and
potentially highly biased) point-source detection rate of $\sim30$\%
within ``classical'' PNe at typical sensitivities of
$10^{29-30}$ erg s$^{-1}$.

For purposes of understanding the potential implications of these
preliminary results for models of magnetic and/or accretion activity
associated wth PN cores, it may be helpful (though potentially
misleading!) to compare PN core X-ray emission levels with those of
solar- and subsolar-mass, pre-main sequence (T Tauri) stars. Evidently
the hard ($\sim$2--8 keV) X-ray emission from such stars is usually coronal
in origin (e.g., Preibisch et al. 2005), although for some fraction of
T Tauri stars the hard X-rays --- as well as some or all of the softer
($\sim$0.3--1.0 keV) emission --- likely arise as a consequence of
magnetically-governed accretion activity (e.g., Kastner et al.\ 2006b;
Telleschi et al.\ 2007). Comparing our {\it Chandra} results for the
X-ray luminosities and luminosity upper limits of point-like PNe X-ray
sources with the $L_X$ distributions of a nearly complete sample of
pre-main sequence stars in Orion (Feigelson et al.\ 2005), one might
speculate that the nondetected PN cores, as a group, are less
magnetically active than $\sim0.5$ $M_\odot$ T Tauri stars; while the
magnetic activity levels of the four relatively X-ray-luminous PN
cores (for which $L_X \sim10^{30-31}$ erg s$^{-1}$) are comparable to
1--3 $M_\odot$ T Tauri stars (ignoring for the moment the likelihood
that the X-ray-luminous cores harbor close binary stars; \S 3.2).

\subsection{Symbiotic Mira systems}

In stark contrast to ``classical'' PNe, it seems that X-ray-luminous
point sources are a common (perhaps ubiquitous) feature of symbiotic
Mira systems: all but one of the six such systems observed thus far by
Chandra --- R Aqr, CH Cyg, Mz 3, Hen 2-104, and Mira itself ---
display X-ray sources at the positions of their central stars.  The
only symbiotic Mira nebula in which no central X-ray point source is
detected, OH 231.8+4.2, possesses a very highly obscured central
region (i.e., a dusty torus). The luminosities of the symbiotic Mira
X-ray point sources range over $\sim4$ orders of magnitude (from
$\sim10^{28}$ to $\sim10^{32}$ erg s$^{-1}$); this emission is
typically harder than that of diffuse (shock-induced) PN X-ray
emission (e.g., Kastner et al.\ 2003) and may be highly variable.

By analogy with low-mass X-ray binaries, the presence of
X-ray point sources within symbiotic Miras is likely indicative of
binary mass transfer processes. Hence, further observations and analyses of 
these X-ray sources should provide unique diagnostics of accretion of AGB
wind material by white dwarf (or, in some cases, main sequence)
secondaries. More generally, studies of compact X-ray sources
within symbiotic Miras should provide insight into the disks, jets,
and disk-jet interactions that are the likely consequence of central
binary systems within PNe (see also Montez \& Kastner, these
proceedings).

\vspace{0.2 cm}
{\it Acknowledgements}: I am indebted to my colleagues Rudy Montez, Bruce
  Balick, and Orsola De Marco for their many contributions to this work.



\end{document}